\documentclass{article}
\usepackage{iclr2025_conference,times}
\usepackage{graphicx}

\usepackage{amsmath,amsfonts,bm}

\def\eqref#1{equation~\ref{#1}}

\def\1{\bm{1}}

\DeclareMathAlphabet{\mathsfit}{\encodingdefault}{\sfdefault}{m}{sl}
\SetMathAlphabet{\mathsfit}{bold}{\encodingdefault}{\sfdefault}{bx}{n}

\usepackage{hyperref}
\usepackage{url}
\newcommand\blfootnote[1]{
  \begingroup
  \renewcommand\thefootnote{}
  \footnote{#1}
  \addtocounter{footnote}{-1}
  \endgroup
}

\title{Therapeutic AI and the Hidden Risks of Over-Disclosure: An Embedded AI-Literacy Framework for Mental Health Privacy}

\author{Soraya S. Anvari \\
Dalhousie University\\
Halifax, Nova Scotia, Canada\\
\texttt{soraya.anvari@dal.ca}
\And
Rina R. Wehbe \\
Dalhousie University\\
Halifax, Nova Scotia, Canada\\
\texttt{rina.wehbe@dal.ca}}
\iclrfinalcopy

\begin{document}

\maketitle
\blfootnote{\textbf{Accepted to SMASH 2025}.}

\begin{abstract}
Large Language Models (LLMs) are increasingly deployed in mental health contexts, from structured therapeutic support tools to informal chat-based well-being assistants. While these systems increase accessibility, scalability, and personalization, their integration into mental health care brings privacy and safety challenges that have not been well-examined. Unlike traditional clinical interactions, LLM-mediated therapy often lacks a clear structure for what information is collected, how it is processed, and how it is stored or reused. Users without clinical guidance may over-disclose personal information, which is sometimes irrelevant to their presenting concern, due to misplaced trust, lack of awareness of data risks, or the conversational design of the system. This overexposure raises privacy concerns and also increases the potential for LLM bias, misinterpretation, and long-term data misuse.
We propose a framework embedding Artificial Intelligence (AI) literacy interventions directly into mental health conversational systems, and outline a study plan to evaluate their impact on disclosure safety, trust, and user experience.
\end{abstract}

\section{Introduction}
In the HCI4Good Lab \footnote{\url{https://hci4good.cs.dal.ca/}}, we design and evaluate digital tools ranging from large displays, mobile games, and AI chatbots to support the mental health education (\cite{10.1145/3715668.3735626,10.1145/3677090, wehbe2022designing}). Through our studies, we have observed two recurring issues when users engage with AI for mental health educational support. First, many users struggle to write effective prompts that generate clear, relevant, and safe responses. Prior work has shown that writing good prompts is difficult (\cite{10.1145/3544548.3581388, KNOTH2024100225}), but this challenge becomes especially critical in sensitive domains such as mental health, where poorly phrased prompts may lead to confusing, inappropriate, or even harmful outputs.
Second, users might overshare personal information, sometimes disclosing sensitive details unrelated to their main concern. This tendency is often reinforced by the conversational style of LLMs, which can create a false sense of privacy and trust.

These challenges are compounded by vague or inaccessible data policies in many mental health applications (\cite{iwaya2022empirical,transparencySurvey2021}). Users are rarely informed about how their conversations are stored, shared, or potentially used for model retraining. Lack of transparency can increase risks such as bias propagation, data leakage, and the misuse of sensitive information \footnote{https://apnews.com/article/betterhelp-ftc-health-data-privacy-befca40bb873661d1f8986bb75d8df07}. Addressing these issues requires both technical safeguards and user-facing education.

\section{Background and Motivation}
Building AI literacy is critical for fostering safer and more informed interactions with conversational systems in mental health contexts. Prior studies demonstrate that AI literacy enhances users’ ability to craft effective prompts and make thoughtful decisions about what information to share, thereby improving both privacy protection and trust in AI systems (\cite{iwaya2022empirical, KNOTH2024100225, 10.1145/3313831.3376727}). Most existing research on AI literacy has focused on general education and skill development, such as helping non-experts understand core AI concepts (\cite{10.1145/3313831.3376727, 10.1145/3544548.3581388}), refining prompt-writing strategies, or making better privacy-related choices (\cite{10.1145/3544548.3581388}). However, in the domain of mental health, work has primarily emphasized identifying risks of disclosure and transparency gaps in current applications (\cite{iwaya2022empirical, transparencySurvey2021}), with little attention to designing interventions that teach users how to engage safely and effectively with AI-based support tools.

Our work is proposing a framework that integrates AI literacy directly into mental health conversational systems, with a specific focus on safe disclosure and prompt literacy. Rather than leaving users to learn through trial and error, we envision AI tools that teach and guide users, using embedded educational strategies and interactive design. This approach moves beyond documenting risks to developing actionable, user-centered methods for safer AI use in sensitive domains.

Our framework can be implemented as an adaptive wrapper layer that interfaces with existing LLM-based systems. This design choice enables compatibility across APIs and model versions while maintaining transparency about educational interventions. The wrapper acts as a local interpretive layer that monitors user inputs and system responses, providing dynamic feedback or prompts when potential oversharing or ambiguous phrasing occurs. 

\section{The Embedded AI Literacy Framework}
We propose a framework that embeds AI literacy directly into mental health educational systems through an adaptive \textbf{wrapper layer} around existing LLMs. This layer operates on the user’s local device or within a secure client environment, monitoring and supporting interactions in real time without modifying or transmitting sensitive data to external servers. By processing inputs and feedback locally, the system minimizes privacy risks while maintaining responsiveness. The framework integrates three literacy principles, each implemented through distinct modules and feedback mechanisms that guide users toward safer and more informed engagement with AI.

\subsection{System Architecture}
\begin{figure}[h]
    \centering
    \includegraphics[width=0.75\textwidth]{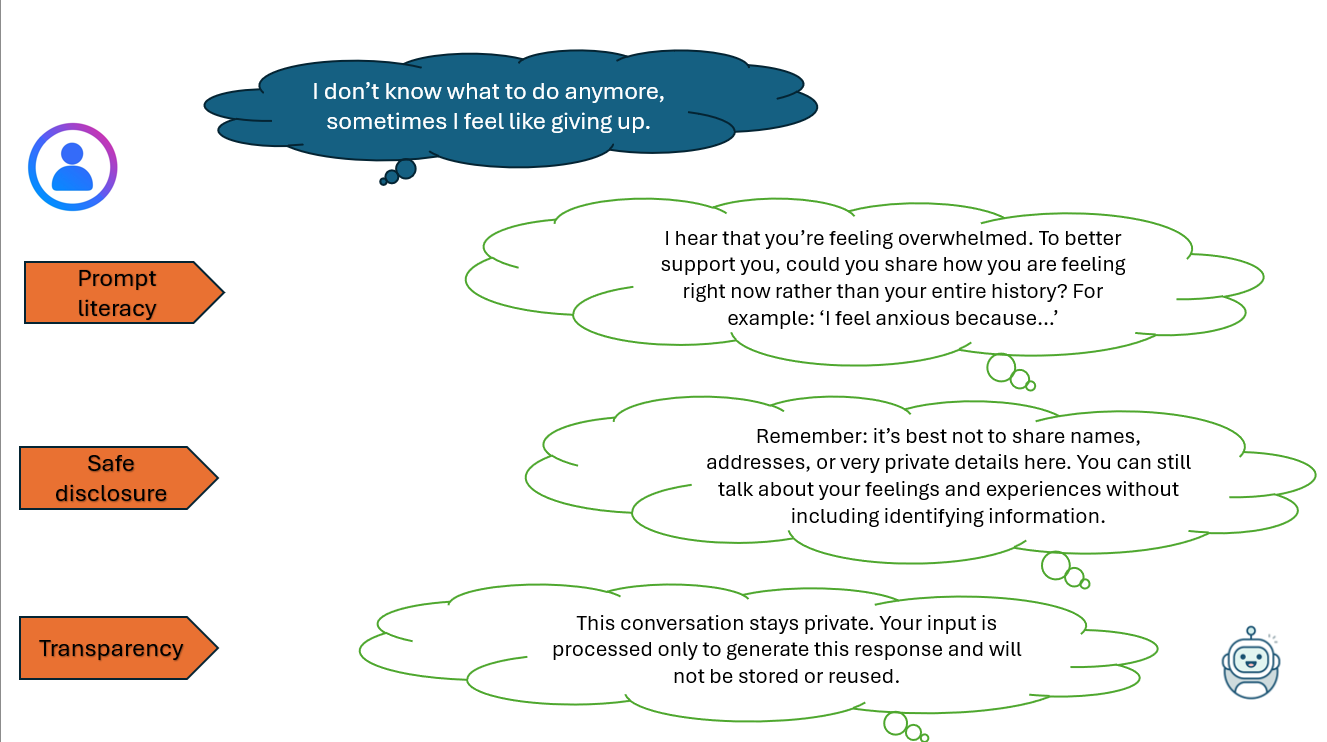}
    \caption{Architecture and flow of the Embedded AI Literacy Framework, showing user interaction, literacy modules (Prompt Coach, Disclosure Monitor, Transparency Engine), and LLM integration.}
    \label{fig:ai-literacy}
\end{figure}

Figure~\ref{fig:ai-literacy} illustrates the core architecture. The framework functions as a middleware layer positioned between the user interface (chat front-end) and the LLM API. Each message passes through the literacy layer before and after model inference. The layer consists of three key modules:

\begin{itemize}
    \item \textbf{Prompt Coach:} Detects vague or ambiguous user inputs and provides structured, example-based reformulations. 
    \item \textbf{Disclosure Monitor:} Classifies user input according to disclosure sensitivity (safe, personal, high-risk) and provides reflection cues. 
    \item \textbf{Transparency Engine:} Generates plain-language explanations about data handling and system behavior, surfaced at relevant conversational moments.
\end{itemize}

Each module is powered by lightweight natural language processing components built on open-source libraries such as \texttt{spaCy} \footnote{\url{https://spacy.io/}}, \texttt{transformers} \footnote{\url{https://huggingface.co/transformers}}, or \texttt{fastText} \footnote{\url{https://fasttext.cc/}}, combined with rule-based pattern detection and prompt-engineered micro-models for classification. Modular design allows local or client-side deployment to preserve privacy.

\subsection{Principle 1: Teach Prompt Literacy}
The \textit{Prompt Coach} uses template-based feedback to teach users how to craft effective prompts. It compares user input against learned patterns of clarity, specificity, and goal alignment. When a prompt is underspecified, the system provides adaptive hints such as:
\begin{quote}
\textit{“Would you like to focus on stress, relationships, or study pressure?”}
\end{quote}
To encourage learning, the system applies dynamic difficulty adjustment: users who demonstrate improved prompt clarity receive subtler guidance, whereas novices receive structured examples or clickable rephrase options. The module can be implemented using a small fine-tuned BERT classifier\footnote{BERT: Bidirectional Encoder Representations from Transformers (\cite{devlin2019bert}), an open-source language model that can be downloaded and fine-tuned locally without transmitting data to external servers.} that rates prompt clarity (1–5) and generates corrective suggestions.

\subsection{Principle 2: Guide Safe Disclosure}
The \textit{Disclosure Monitor} combines named entity recognition (NER) and semantic classification to detect potentially sensitive or identifiable information (e.g., names, locations, detailed life events). A disclosure taxonomy distinguishes:
\begin{itemize}
    \item \textit{Safe:} general feelings or reflections (“I felt anxious today.”)
    \item \textit{Personal:} identifiable but non-critical details (“My friend Sarah…”)
    \item \textit{High-risk:} potentially harmful or crisis-related content
\end{itemize}
When a personal or high-risk disclosure is detected, the interface displays a prompt such as:
\begin{quote}
\textit{“This message may include personal details. Would you like to rephrase or continue?”}
\end{quote}
High-risk cases automatically trigger referral links (e.g., national help lines or campus resources). This classification pipeline can be implemented using pre-trained NER via \texttt{spaCy} (\cite{spacy2015}), combined with a lightweight fine-tuned transformer model such as DistilBERT (\cite{sanh2019distilbert}) or RoBERTa-base (\cite{liu2019roberta}) for disclosure intent classification. All components can be deployed locally, ensuring that sensitive user text remains on the client device throughout analysis.

\subsection{Principle 3: Maintain Trust Through Transparency}
The framework helps users understand how the system handles their data by providing explanations during the conversation.
These explanations appear when relevant, such as when users ask about privacy or when sensitive topics arise, so that they inform without interrupting the flow of conversation. The transparency messages clearly state what data is collected, how it is used, and what is not stored, using simple, non-technical language. 
This approach builds trust gradually, helping users feel informed and in control without overwhelming them with technical details.

\section{Evaluation and Research Plan}
The evaluation focuses on non-clinical users who interact with mental health chatbots for reflection and educational purposes. The framework will be co-developed with health and mental health experts to ensure that literacy interventions support safe self-expression without restricting meaningful dialogue.

A longitudinal, repeated-measures user study can be conducted to compare two chatbot conditions: (1) a baseline system without literacy features and (2) a version with the embedded AI literacy layer. Participants may engage with both systems across multiple sessions over several weeks to observe changes in prompting behavior, disclosure patterns, and trust development over time. Each session could include short reflective chat tasks (10–15 minutes) followed by brief post-session surveys, and a final interview at the end of the study. All interactions would be anonymized and conducted under approved research ethics protocols.

\subsection{Measures}
Evaluation can focus on three key dimensions aligned with the framework: \textit{prompt literacy}, \textit{safe disclosure}, and \textit{trust and transparency}.

Prompt literacy can be examined through changes in prompt clarity and user reflection. Chat transcripts may be coded using a clarity rubric adapted from prior AI literacy studies, complemented by self-reported learning items (e.g., “I learned how to ask clearer questions to the chatbot”).

Safe disclosure can be measured by analyzing how often participants share personal or high-risk details. Each conversation could be annotated into safe, personal, and high-risk categories to calculate disclosure proportions, following prior work on privacy and disclosure in mental health technologies (\cite{iwaya2022empirical, bhatia2021mental, wehbe2022designing}). Perceived privacy and safety can be captured using selected items from the Internet Users’ Information Privacy Concerns (IUIPC) scale (\cite{malhotra2004iuipc}). Trust and transparency can be evaluated through the Trust of Automated Systems Test (TOAST) (\cite{wojton2020toast}) and short comprehension questions about how the system handles user data. Follow-up interviews may explore users’ emotional comfort, system credibility, and understanding of transparency cues.

Usability and engagement could also be assessed using the System Usability Scale (SUS) to capture participants’ overall experience. Quantitative data could be analyzed using within-subject comparisons, while qualitative insights from interviews and transcripts may undergo thematic analysis.

This evaluation approach enables examination of how literacy-embedded chatbots might help users phrase clearer prompts, reduce unsafe disclosure, and enhance understanding of data practices, ultimately supporting greater perceived trust and safety in AI-supported mental health education. Insights from such studies can inform future refinements of the framework and collaborations with clinicians and AI-HCI researchers.

Attending SMASH 2025 provides an opportunity to refine and expand this work in dialogue with experts across computer science, healthcare, and the social sciences. Collaboration with peers at the symposium will help us strengthen our research design, explore interdisciplinary partnerships (particularly with clinicians and ethicists), and identify best practices for accountability in mental health AI. We see this conference as a great chance to connect with collaborators who share our goal of developing responsible, transparent, and user-centered AI for mental health support.

\bibliography{iclr2025_conference}
\bibliographystyle{iclr2025_conference}

\end{document}